\documentclass[aps,twocolumn,pr,superscriptaddress,longbibliography]{revtex4-1}
\usepackage{times,amsmath,graphicx}
\usepackage[colorlinks=true,citecolor=blue,urlcolor=blue,anchorcolor=blue,linkcolor=blue]{hyperref}
\usepackage{amssymb}
\usepackage{multirow}
\usepackage{color}

\begin{document}
\title{Multi-gap and high-$T_{c}$ superconductivity in metal-atom-free borocarbides: Effects of dimensional confinement and strain engineering}

\author{Hao-Dong Liu}
\affiliation{Institute of Applied Physics and Computational Mathematics and National Key Laboratory of Computational Physics, Beijing 100088, China}
\affiliation{Graduate School, China Academy of Engineering Physics, Beijing 100088, China}

\author{Wei-Yi Zhang}
\affiliation{School of Manufacturing Science and Engineering, Southwest University of Science and technology, MianYang 621010, China}

\author{Zhen-Guo Fu}
\thanks{E-mail: fu\_zhenguo@iapcm.ac.cn}
\affiliation{Institute of Applied Physics and Computational Mathematics and National Key Laboratory of Computational Physics, Beijing 100088, China}
\affiliation{Graduate School, China Academy of Engineering Physics, Beijing 100088, China}

\author{Bao-Tian Wang}
\affiliation{Institute of High Energy Physics, Chinese Academy of Sciences, Beijing 100049, China}
\affiliation{Spallation Neutron Source Science Center, Dongguan 523803, China}

\author{Hong-Yan Lu}
\affiliation{School of Physics and Physical Engineering, Qufu Normal University, Qufu 273165, China}

\author{Hua-Jie Song}
\thanks{E-mail: song\_huajie@iapcm.ac.cn}
\affiliation{Institute of Applied Physics and Computational Mathematics and National Key Laboratory of Computational Physics, Beijing 100088, China}

\author{Ning Hao}
\affiliation{Anhui Province Key Laboratory of Low-Energy Quantum Materials and Devices, High Magnetic Field Laboratory, HFIPS, Chinese Academy of Sciences, Hefei, Anhui 230031, China}

\author{Ping Zhang}
\thanks{E-mail: zhang\_ping@iapcm.ac.cn}
\affiliation{Institute of Applied Physics and Computational Mathematics and National Key Laboratory of Computational Physics, Beijing 100088, China}
\affiliation{Graduate School, China Academy of Engineering Physics, Beijing 100088, China}
\affiliation{School of Physics and Physical Engineering, Qufu Normal University, Qufu 273165, China}

\begin{abstract}
Pure borocarbides suffer from limited superconducting potential due to intrinsic structural instability, requiring transition/alkali metals as dual-functional stabilizers and dopants. Here, by combining high-throughput screening with anisotropic Migdal-Eliashberg (aME) theory, we identify dynamically stable borocarbides where high-$T_{c}$ superconductivity predominately originates from $E$ symmetry-selective electron-phonon coupling (EPC). The six distinct superconducting gaps emerge from a staircase distribution or uncoupling of EPC strength $\rho(\lambda^{el}_{k})$ across each Fermi surface (FS) sheet, constituting a metal-free system with such high gap multiplicity. Crucially, dimensional reduction from bulk to surface strengthens $E$-symmetry EPC and enhances $T_{c}$ from 32 K (3D bulk) to 75 K (2D surface), a result that highlights structural confinement as a key design strategy for observing high $T_{c}$. External strain further optimizes the competition between EPC strength and characteristic phonon frequency to achieve $T_{c} >90$ K. This work reveals a systematic correlation between structural dimensionality and gap multiplicity and establishes borocarbide as a tunable platform to engineer both high-$T_{c}$ and multi-gap superconductivity.

\end{abstract}
\maketitle

$Introduction$---The century-long quest for high-temperature superconductivity since its discovery in 1911 has developed fundamental paradigms in condensed matter physics. A pivotal advance emerges with MgB$_{2}$ \cite{Nagamatsu2001}, whose superconductivity with $39$ K and two-gap behavior \cite{Choi2002,PhysRevB.92.054516} demonstrate the potential of phonon-mediated pairing in layered hexagonal systems. Subsequent studies focus on isostructural diborides where Mg is substituted with transition \cite{10.1093/nsr/nwad034,Lim2022,PhysRevB.64.224509,PhysRevB.64.140509,PhysRevB.76.214510} and alkali \cite{PhysRevMaterials.6.024803,PhysRevB.80.064503,PhysRevB.64.020502,PhysRevB.64.224509,PhysRevB.64.140509} metals, followed by systematic exploration of partial C substitution in B layers, aiming to enhance EPC \cite{PhysRevLett.95.237002,Weller2005,PhysRevLett.95.087003,PhysRevB.108.214507}. Despite these efforts, the majority of derivatives exhibit $T_{c}$ values constrained by the McMillan limit ($39$ K), reflecting inherent limitations in optimizing competing parameters (coupling strength $\lambda$ vs. phonon frequency $\omega_{log}$) within conventional pairing frameworks.

\begin{figure*}[htbp!]
	\centering
	\includegraphics[width=0.7\linewidth]{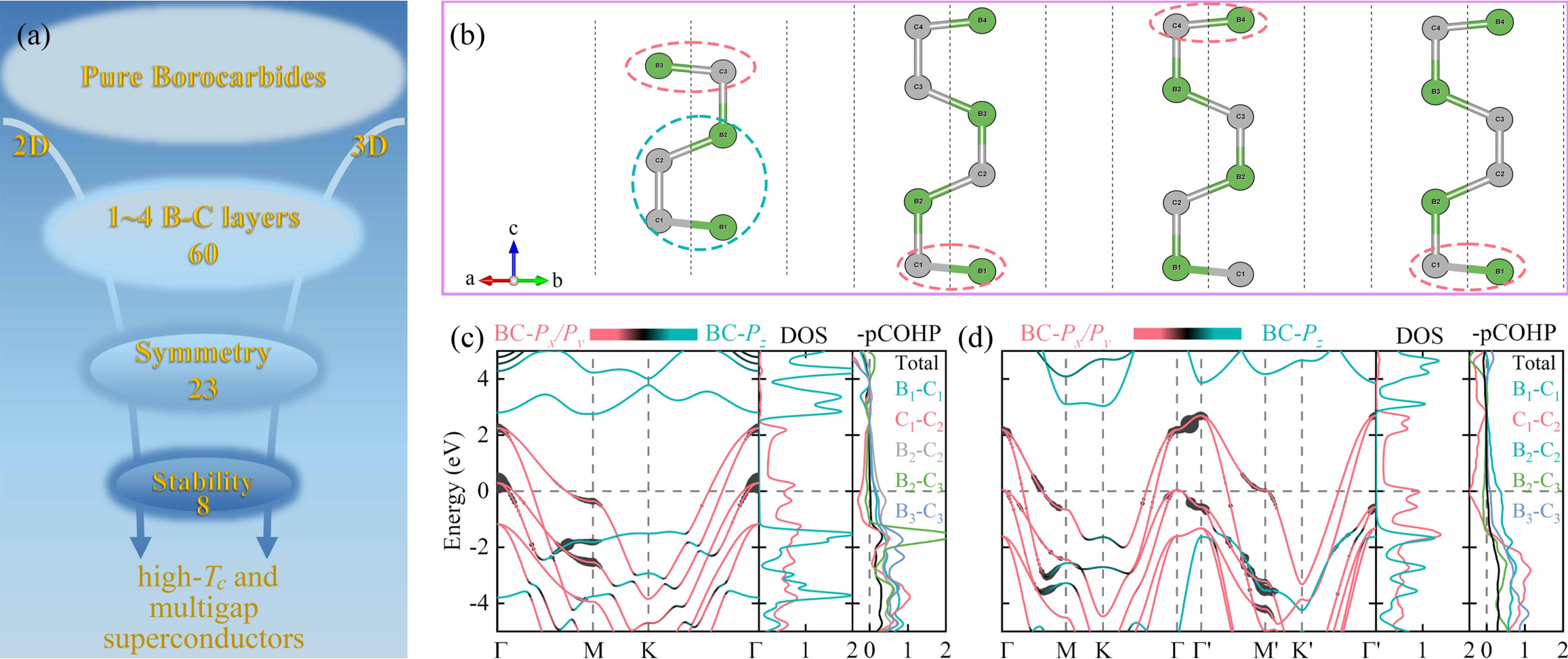}
	\caption{(a) Schematic diagram of the high-throughput screening procedure for computationally designing pure borocarbides. Panel (b) shows crystal structures of the AAB, ABAA, ABAB, and ABBA configurations within a unit cell, arranged from left to right. Green and gray balls represent boron and carbon atoms, respectively. The atoms contributing to superconductivity are marked with dashed lines. Comparison of the electronic structure, DOS, and -pCOHP of the AAB configuration in (c) 2D and (d) 3D cases.}
	\label{fig:fig}
\end{figure*}

Recent advancements in superconducting materials design highlight dimensional control across several complementary strategies. Three-dimensional (3D) layered borocarbides achieve enhanced B-C covalent bonding through controlled carbon substitution, where optimized orbital hybridization balances EPC strength ($\lambda$) and lattice stability \cite{RIBEIRO2003227,PhysRevB.64.134513,PhysRevLett.88.127001,PhysRevB.64.140509,PhysRevB.91.045132,doi:10.1021/acs.inorgchem.9b02709,PhysRevB.102.144504,PhysRevB.107.134502,PhysRevB.107.014508,C8CP07634K,D4TC00210E,tomassetti2024,PhysRevB.108.024512}. Simultaneously advancing in lower dimensions, zero-dimensional (0D) clathrate frameworks utilize cage confinement effects to strengthen pairing interactions in boron-carbon clusters, bypassing the electron doping bottlenecks inherent to bulk systems \cite{PhysRevB.103.144515,PhysRevB.105.064516,PhysRevB.105.224514,ZHANG2024129643,PhysRevB.109.184517,PhysRevB.109.144509}. Parallel progress in two-dimensional (2D) systems employs surface engineering of atomically thin borocarbides \cite{JISHI2003358,PhysRevB.101.094501,PhysRevB.104.054504,Singh2022,doi:10.1021/acs.nanolett.2c05038,LiBCH,PhysRevB.111.184502}, where dimensional reduction enables collaborative optimization of both $\lambda$ and $\omega_{log}$, overcoming the conventional trade-off observed in isotropic Eliashberg theory and realizing 2D multi-gap superconductors \cite{GINZBURG1964101,PLUMMER2003251,PhysRevLett.21.1320,PhysRevB.83.220503,PhysRevB.87.140503,Liu2022}. This multi-dimensional paradigm demonstrates how structural confinement effects (3D $\rightarrow$ 0D $\rightarrow$ 2D) as well as B-C bonding configurations therein permit independent manipulation of competing superconducting parameters.

Despite these advances, a critical unresolved issue persists: stoichiometric B-C compounds lacking metal constituents remain unexplored theoretically and experimentally. We address this by systematically investigating pure borocarbides across 2D and 3D configurations. First-principle calculations identify trilayer system as a dynamically stable minimum unit, establishing the foundation for analyzing their superconducting mechanisms. Our investigations aim to elucidate the stable mechanism and demonstrate the origin of superconductivity in these structures, with particular interest in their potential for high $T_{c}$, multi-gap behavior, and the dimensional or straining manipulation on superconductivity. These efforts resolve the long-standing paradox between structural stability and strong EPC in metal-free borocarbides systems.

$Screening$---To identify a robust materials design framework, we develop a high-throughput screening protocol [Fig. 1(a)] that systematically evaluates pure borocarbides possible configurations (see Section II of Supplementary Materials (SM) \cite{SM}). Following initial symmetry constraints and structural uniqueness filtering of 60 candidate structures, 23 configurations are retained for further analysis. Through rigorous dynamic stability assessments based on full phonon dispersion criteria, 8 structures are validated as viable candidates. Detailed computational methodology is provided in Section I of SM \cite{SM}. The identified structures are subsequently subjected to systematically stable validation: (1) Thermodynamic stability through formation and cohesive energy analyses, (2) dynamic stability via phonon dispersion examination, and (3) mechanical stability through linear elastic constants evaluation. Comprehensive validations are shown in Section III of SM \cite{SM}. 

\begin{figure*}[htbp!]
	\centering
	\includegraphics[width=1.\linewidth]{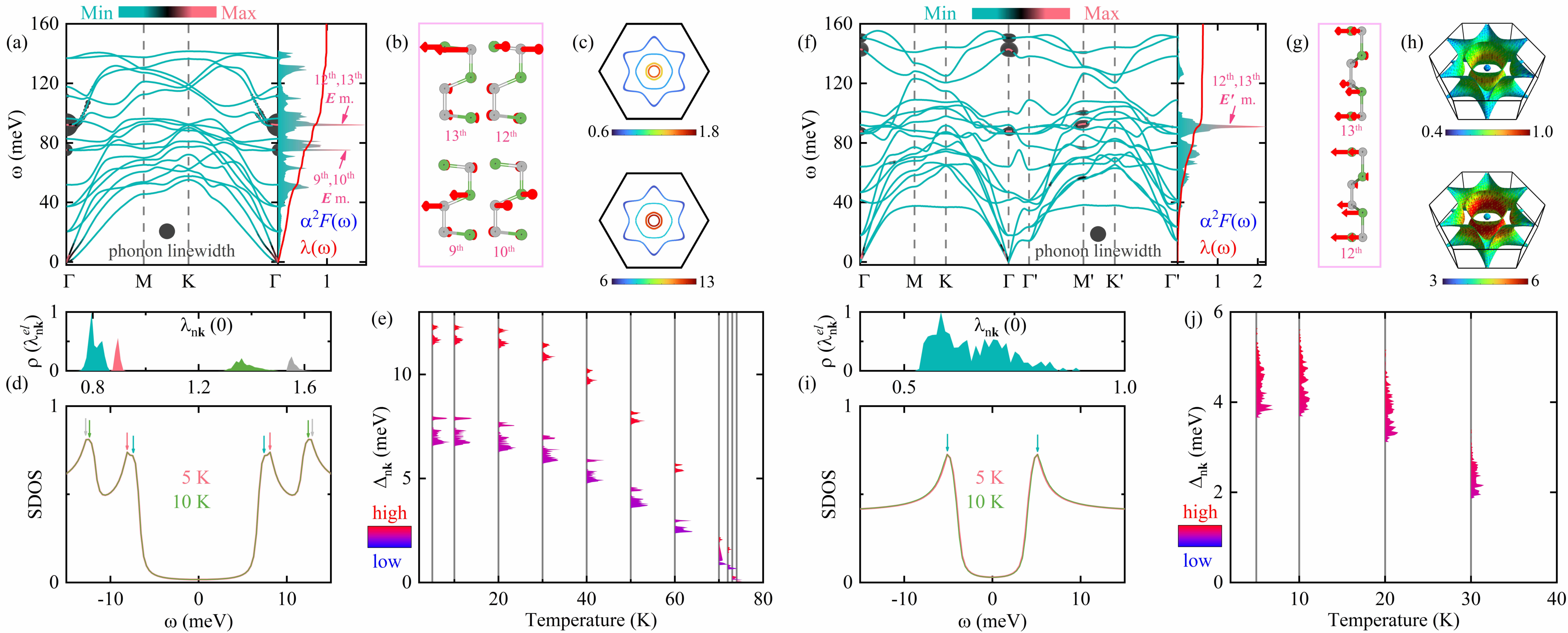}
	\caption{Phonon and superconductivity in 2D (a) - (e) and 3D (f) - (j) AAB configurations. (a) Phonon dispersion weighted by $\lambda^{\mathit{ph}}_{\mathbf{q}}$ (left) and isotropic $\alpha^{2}F(\omega)$ and cumulative frequency-dependent EPC strength $\lambda(\omega)$ (right). The color bar, ranging from blue to red, represents the shift from minimum to maximum values of $\lambda^{\mathit{ph}}_{\mathbf{q}}$. m. is an abbreviation for modes. (b) Atomic vibrational modes at $\Gamma$ point corresponding to the main peaks in the $\alpha^{2}F(\omega)$. Red arrows represent vibrational direction. (c) EPC parameters $\lambda^{\mathit{el}}_{\mathbf{k}}$ (top) and momentum-resolved superconducting gap $\Delta_{n\mathbf{k}}(\omega=0)$ at $T$=5K (bottom) are projected onto the FS. (d) Strength distribution $\rho (\lambda^{\mathit{el}}_{\mathbf{k}})$ of $\lambda^{\mathit{el}}_{\mathbf{k}}$ (top) and SDOS at $T$ = 5 and 10 K (bottom). The arrows indicating SDOS peaks are color-matched to the strength distribution $\rho (\lambda^{\mathit{el}}_{\mathbf{k}})$ shown in top panel. (e) Energy distribution of $\Delta_{n\mathbf{k}}$ versus $T$. (f)and (j) The corresponding phononic and superconducting results for 3D AAB configuration.}
\end{figure*}


The identified stable configurations exhibit distinct structural fingerprints that directly correlate with their superconducting performance. As detailed in Section IV of the SM \cite{SM}, critical parameters including lattice constants, space/point groups, bond lengths, and Wyckoff positions establish fundamental structure-property relationships. Distinct from ABAA/ABAB/ABBA configurations, our calculations reveal a characteristic 3D-to-2D dimensional reconstruction phenomenon in AAB configuration, where the naming details are presented in the Section II of SM \cite{SM}. Explicitly, 3D-periodic AAB configuration possesses planar B$_{3}$/C$_{3}$ [Fig. S7(a)], contrasting with the rippled 2D counterpart ($\delta z=0.08$ {\AA} wrinkle). This structural transition elevates symmetry from $C_{3v}^{1}$ (2D) to $C_{3h}^{1}$ (3D) through mirror plane recovery, and creates degenerate electronic states at $E_{F}$. However, the following calculations reveal that such symmetry-enforced degeneracy paradoxically suppresses multi-gap features, ultimately reducing $T_{c}$ by $57\%$ compared to symmetry-broken configurations (see the following text).

$Electronic$ $properties$---Bader \cite{HENKELMAN2006354} charge analysis (Table S6) reveals systematic electron transfer from boron to carbon atoms across all configurations, showing pronounced electron localization at edge atoms in each configuration. This spatial selectivity aligns with established mechanisms of surface-enhanced superconductivity \cite{doi:10.1021/acs.nanolett.2c05038} and is quantitatively supported by atomic volume fraction analysis. Charge density difference (Fig. S7) demonstrates two concurrent electronic reorganization processes: depletion of out-of-plane $p_{z}$ orbitals and enhancement of in-plane $p_{x}$/$p_{y}$ hybridization for both atomic species. Electrons seem to favorably transfer to B-C bond rather than C-C bond. The resulting electron accumulation at B-C bonding regions, visualized through electron localization function (ELF, Fig. S8) analysis, creates an optimal electronic environment for strong EPC. Moreover, dimensional confinement in 2D configurations may amplify these effects through the enhanced charge localization at structural edges and reduced dielectric screening of Coulomb interactions. These combined factors critically contribute to the high-$T_{c}$ superconductivity. The observed orbital reconstruction mechanism, which couples directional hybridization with surface-dominated charge distribution, provides fundamental electronic basis for understanding the multi-gap superconducting behavior.

The calculated electronic band structures and density of states (DOS) presented in Figs. 1(c)-1(d) and Fig. S9, with corresponding Brillouin zones (BZ) and high-symmetry routes/points annotated in Fig. S10, reveal critical features underlying the superconducting mechanism for 2D and 3D configurations. One can see that dominant contributions stem from B-C $p_{x}$ and $p_{y}$ orbitals near the Fermi level (FL), consistent with the aforementioned charge transfer analysis, while $s$- and $p_{z}$-orbital participation remains negligible. Both 2D and 3D configurations exhibit metallic behavior characterized by multiple bands crossing the FL, whose the number actually corresponds to ($N_{atom}$ - $2$). Enhanced EPC potential emerges from the pronounced electrons, which are marked with dark gray shadows (electron linewidths \cite{Giustino} [Eq. (S9)]) concentrated at the $\Gamma$-point and saddle points near $M$ ($M^{'}$). Notably, dimensional increasing from 2D to 3D AAB configurations induces a characteristic FL shift to higher DOS regions, accompanied by symmetry restoration. Moreover, for these 8 systems, the several bands crossing the FL at $\Gamma$ point are assigned the $E$-symmetry constraints. Detailed orbital-projected and 3D band structures are provided in Figs. S11-S18. Near-identical $\Gamma$-$M$-$K$-$\Gamma$ and $\Gamma'$-$M'$-$K'$-$\Gamma'$ dispersion relations and corresponding antom-projected DOS analyses (Figs. S12-S14 and S16-S18) confirm the preservation of quasi-2D characteristics in other ABAA/ABAB/ABBA configurations, further demonstrated by FS [Figs. S19(g)-S19(x)]. Additionally, FS exhibit strong anisotropy in both velocity distribution and orbital composition, establishing the foundation for applying aME theory \cite{migdal1958interaction,eliashberg1960interactions,Philip} to unravel the superconductivity mechanisms.

To characterize interatomic bonding interactions, we calculated the negative projected crystal orbital Hamiltonian population (-pCOHP) \cite{doi:10.1021/j100135a014,COHP} using the LOBSTER code \cite{doi:10.1021/jp202489s}, where positive and negative values correspond to bonding and antibonding states, respectively. As shown in Figs. 1 and S10, all configurations exhibit positive total -pCOHP values, confirming their structural stability through dominant bonding contributions. A notable dichotomy emerges in the $c$-axis atomic interactions, which display antibonding (negative values) character near the FL opposite to other planar bonding behaviors. This directional anisotropy reveals preferential occupation of antibonding states by $p_{z}$-orbitals along the $c$-axis, while $p_{x}$/$p_{y}$ orbitals maintain covalent bonding in planar. Such orbital-selective bonding configurations create favorable conditions for strong EPC, as the partially filled bonding states near FL enable enhanced electron-lattice interactions while maintaining structural integrity through in-plane covalent bonds \cite{PhysRevLett.86.4366,PhysRevB.107.134502,PhysRevB.110.064513}. This bonding hierarchy suggests critical insights into the relationship between electronic structures and superconducting property discussed subsequently.

$Phonon$ $properties$--To elucidate the phonon-mediated pairing mechanism, we systematically analyze vibrational properties and their coupling contributions in these configurations (see Fig. 2 and subsection A of Section VI in SM \cite{SM}). Figure 2 reveals distinct dimensional effects of AAB configuration on phonon dispersion characteristics. One can see that, compared to the 2D counterpart, the 3D configuration exhibits enlarged phonon spectra with a $\sim10\%$ increase in maximum frequency and $\sim20\%$ enhancement of $\omega_{log}$ [Eq. (S5)]. Atomic vibrational analysis (Figs. S21 and S25) demonstrates complementary motion patterns: B/C in-plane vibrations span the full frequency spectrum, while out-of-plane modes concentrate within the first acoustic branch and medium-frequency region ($40$ - $120$ meV). In particular, 3D AAB structure shows prominent flat-band phonon ($\sim37$ meV) from B$_{3}$-atom $z$-direction vibration that generates singularly large  phonon DOS (PhDOS) [see Fig. S25(c)]. Overall, the mass similarity between B and C atoms produces comparable PhDOS distributions across all configurations, though dimensional restructuring modifies mode participation in EPC. Phononic vibrational analysis of additional ABAA/ABAB/ABBA configurations presented in subsection B of Section VII in SM \cite{SM} will not be discussed in detail for briefness.

Furthermore, our analysis identifies dimension-dependent EPC mechanisms through comparative examination of vibrational modes. As shown in Figs. 2(a)-(b) and 2(f)-(g), the 2D AAB configuration exhibits four EPC-active optical branches ($9^{th}$, $10^{th}$, $12^{th}$, and $13^{th}$ modes) compared to two dominant modes ($12^{th}$ and $13^{th}$) in its 3D counterpart, as evidenced by $\Gamma$-point phonon linewidth accumulation [Eq. (S1)] and distinct peaks in the Eliashberg function $\alpha^{2}F(\omega)$, where the 2D configuration shows double-peak spectral features while a single broad peak is observed in 3D case. Symmetry decomposition reveals these coupling channels derive from $E$- and $E'$-symmetry phonon modes for 2D and 3D cases, respectively, combining infrared and Raman (I+R) activities. Crucially, the active phonon modes in both dimensionalities originate from in-plane B-C atomic vibrations, establishing a direct correlation between planar lattice dynamics and superconductivity. The preserved dominance of $12^{th}$/$13^{th}$ modes across different dimensionality suggests robust EPC channels rooted in planar B-C bonding networks, providing microscopic insight into structure-property relationships for superconducting design.

$Superconductivity$---The dimensional reduction from 3D to 2D systems fundamentally modifies the EPC landscape through structural confinement effects, which is demonstrated by the density functional perturbation theory within QUANTUM ESPRESSO (QE) package \cite{Paolo} combined with aME theory and tight-binding models (Fig. S29) in EPW code \cite{PhysRevB.87.024505,Giustino,Jesse}. Explicitly, as shown in Figs. 2(a) and 2(f), the 2D AAB configuration exhibits enhanced EPC strength ($\lambda = 0.99$) compared to its 3D counterpart ($\lambda = 0.63$), while simultaneously softening $\omega_{log}$ through surface-induced vibrational modes. This enhancing EPC $\lambda$ drives a remarkable $T_{c}$ enhancement from $32$ K in the 3D bulk to $75$ K in the 2D surface system [see Figs. 2(e) and 2(j)], while showing progressive gap reduction down to zero with increasing temperature. Aforementioned orbital-resolved analysis demonstrates that dimensional confinement selectively strengthens $p_{x}$/$p_{y}$ orbital hybridization near the FL while promoting $E$-symmetry selective coupling between in-plane B/C vibrations and electronic states, unlike $E^{'}$ symmetry in 3D case, which is responsible for the increase of $T_{c}$ value. On contrast, for other configurations, due to the quasi-2D characteristic, the EPC is still dominated by $E$-symmetry phonon modes, resulting in negligible change in $T_{c}$ (subsection B of Section VII in SM \cite{SM}). Consequently, coupling between electron and phonon both with $E$ symmetry can produce an additional route for designing high-$T_{c}$ superconductors.

Furthermore, the above and detailed analysis in subsection B of Section VII in SM \cite{SM} indicates that surface atoms circled by the dashed lines in Fig. 1(b) rather than the inner layers predominantly contribute to high $T_{c}$. Consistent with prior reports \cite{GINZBURG1964101,PhysRevB.7.3028,PLUMMER2003251,PhysRevLett.21.1320,PhysRevB.83.220503,PhysRevB.87.140503,PhysRevX.11.021065,Liu2022,doi:10.1021/acs.nanolett.2c05038}, the critical role of surface atoms in high-$T_{c}$ superconductivity aligns with established confinement effects in low-dimensional systems, where optimized phonon softening overcomes conventional bulk material limitations. Our findings establish structural dimensionality as a key design parameter for simultaneously enhancing EPC and producing multi-gap superconductivity in a case study of these systems. Moreover, the 2D configuration develops four distinct superconducting gaps [Fig. 2(d)], contrasting with the single-gap behavior observed in the 3D counterpart [Fig. 2(i)], which are evidenced by superconducting density of states (SDOS) and $\rho (\lambda^{\mathit{el}}_{\mathbf{k}})$. This multi-gap superconductivity originates from the significant disparity in the distribution of EPC strength across each FS sheet as discussed following.

\begin{figure}[htbp!]
	\centering
	\includegraphics[width=0.8\linewidth]{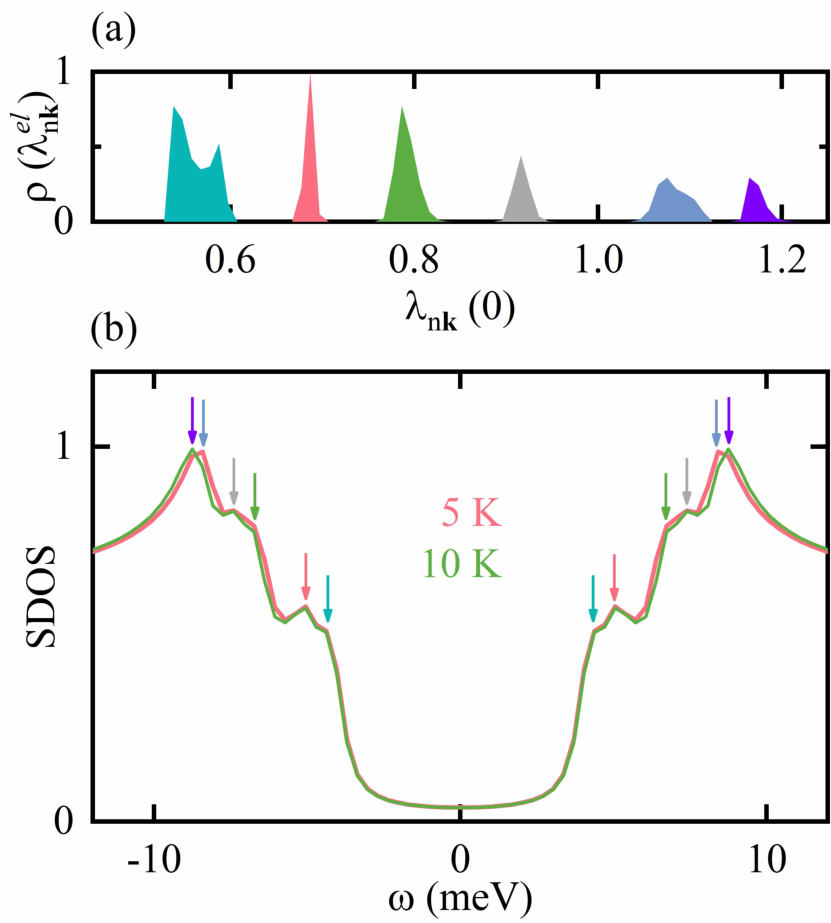}
	\caption{(a) Strength distribution $\rho (\lambda^{\mathit{el}}_{\mathbf{k}})$ of EPC parameters $\lambda^{\mathit{el}}_{\mathbf{k}}$ (top) and (b) SDOS at $T$ = 5 and 10 K (bottom) for the 2D ABBA configuration. The arrows indicating SDOS peaks are color-matched to the strength distribution $\rho (\lambda^{\mathit{el}}_{\mathbf{k}})$ shown in panel (a).}
\end{figure}

The emergence of multi-gap superconductivity carries fundamental significance, and present intriguing research frontier. Unlike conventional two-gap systems, configurations with more gap multiplicities enable exploration of competing quantum orders and topological excitations, as evidenced by studies of time-reversal symmetry breaking, fractional vortices, and hidden criticality \cite{PhysRevB.81.134522,Milosevic_2015,PhysRevB.87.134510,PhysRevLett.89.067001,PhysRevLett.107.197001,PhysRevLett.108.207002,
PhysRevB.85.134514,daSilva2015}. Our identification of four-gap superconductivity in the 2D AAB configuration and six-gap behavior in ABBA borocarbides (Fig. 3) may provide unprecedented experimental targets for these phenomena. Consistent with the physical mechanism of the four superconductivity gaps in 2D ABB configuration, momentum-resolved analysis reveals that the six distinct gaps in ABBA configuration originate from a staircase distribution or uncoupling of EPC strength $\rho(\lambda^{el}_{k})$ across each FS sheet [Fig. 32(c)], unlike strong coupling in two outer FS sheets of 3D ABB configuration [Fig. 2(h)], where dimensional confinement suppresses interband scattering while maintaining anisotropic pairing interactions. This mechanism demonstrates a novel pathway to simultaneously enhance critical temperatures and gap multiplicity in superconductors, achieved through dimensionally controlled FS engineering that preserves strong pairing interactions while suppressing interband decoherence. Furthermore, the demonstrated tunability of gap multiplicity (from $3$ to $6$) across different configurations (see Figs. S30-S32) establishes borocarbides as a unique materials platform, where structural dimensionality directly controls both superconducting transition temperatures and quantum complexity. Our results demonstrate a dimensional control methodology that enables selective gap engineering in superconducting materials.

\begin{figure}[htbp!]
	\centering
	\includegraphics[width=0.8\linewidth]{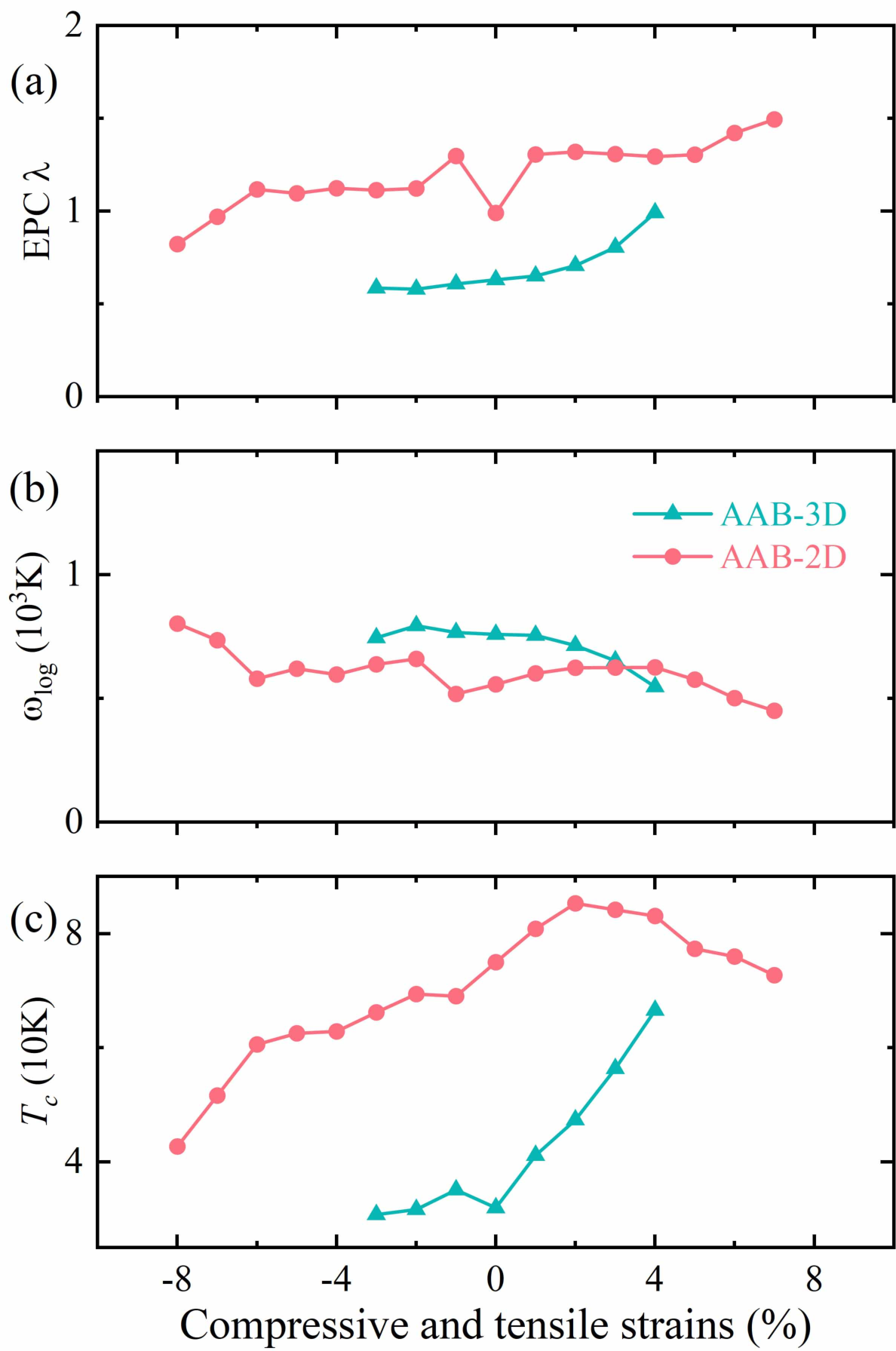}
	\caption{The superconducting parameters (a) $\lambda$, (b) $\omega_{log}$, and (c) $T_{c}$ under various values of strains.}
\end{figure}

$Strains$---Here, we further investigate straining effects on superconducting parameters ($\lambda$, $\omega_{log}$, and $T_{c}$) by applying compressive and tensile strains through $\varepsilon= (a - a_{0})/a_{0}\times100\%$ along each crystallographic axis until structural instability occurs. Our analysis reveals a fundamental competition \cite{PhysRevLett.132.166002} between $\lambda$ and $\omega_{log}$ as follows. Due to the difference introduced by dimensional change in AAB configurations, which are discrepant with the others' quasi-2D characteristic, only the configurations are present in Fig. 4 to further discuss and others are shown in SM \cite{SM}. As shown in Figs. 4(a) and 4(b), it is clear that compressive strain generally suppresses $\lambda$ while enhancing $\omega_{log}$, whereas tensile strain produces the opposite trend. In Fig. 4(c), $T_{c}$ demonstrates non-monotonic dependence governed by the dominant parameters evolution. Specifically, $T_{c}$ increases when $\lambda$ enhancement outweighs the suppression included by the decreasing of $\omega_{log}$, but decreases under dominant $\omega_{log}$ reduction. We can see that dimensional confinement amplifies this tunability with 2D-AAB configuration exhibiting greater $T_{c}$ enhancement (up to over $90$ K) than their 3D-AAB counterpart under equivalent strain conditions. This dimensional sensitivity originates from surface-induced phonon softening and optimized $E$-symmetry selective coupling in reduced geometries. Therefore, strain engineering is an effective strategy for collaborative optimization of competing superconducting parameters, while revealing dimensional reduction as a critical factor in overcoming conventional McMillan limit constraints through synergistic $\lambda-\omega_{log}$ manipulation.

In summary, we presented a comprehensive theoretical investigation of pure borocarbides that integrates high-throughput computational screening with advanced many-body analysis to identify and characterize novel superconducting systems. Our study established pure borocarbides as a unique class of metal-free superconductors exhibiting unprecedented multi-gap superconductivity, including a prototypical six-gap system, with critical temperatures approaching liquid nitrogen temperature. The emergent multi-gap behavior originates from a staircase distribution or uncoupling of EPC strength $\rho(\lambda^{el}_{k})$ across each FS sheet. Crucially, dimensional reduction from 3D bulk to 2D surface dramatically enhances $T_{c}$ while producing gap multiplicity, demonstrating structural confinement as a key design principle for reconciling high-$T_{c}$ and multi-gap characteristics. Furthermore, we observe that the pristine high $T_{c}$ approaching $77$ K can be enhanced up to over $90$ K through optimized strain engineering, which induces a tunable competition between EPC coupling strength and characteristic phonon frequency. Our findings establish borocarbides as a unique materials platform for realizing high-$T_{c}$ and multi-gap superconductivity, while providing a universal design framework for engineering quantum materials through dimensional control and lattice-strain manipulation.

$Acknowledgments$---This work was supported by the National Natural Science Foundation of China (Grant Nos. 12022413, 12175023, 12074213, 11574108, 12275031, 92265104) and the National Key R\&D Program of China (Nos. 2022YFA1403100, 2022YFA1403200, 2024YFA1613200).

\bibliography{BC.bib}
\end{document}